%% file: main.tex
\documentclass[times,10pt,twocolumn]{article}   % icde style
\usepackage{latex8}  % icde style
\usepackage{times}  % icde style

\newcommand{\DrawLines}[1]{}
\renewcommand{\DrawLines}[1]{#1}

\usepackage{amsmath}
\usepackage{amsfonts}
\usepackage{pstricks,graphics,graphicx}
\DrawLines{
\usepackage{pstcol,pst-char,pst-text,pst-plot,pst-node}
}
\usepackage{algorithm}
\usepackage{algorithmic}
\usepackage{epsfig}

\usepackage{atbeginend}
\AfterBegin{itemize}{\addtolength{\itemsep}{-6pt}} 
\AfterBegin{enumerate}{\addtolength{\itemsep}{-3pt}} 
\newcommand{\BlockComment}[1]{}

\newcommand{\IfTechReport}[2]{#2}  % non tech report version

\newtheorem{theorem}{Theorem}
\newtheorem{lemma}[theorem]{Lemma}

\newtheorem{definition}[theorem]{Definition}

\newcommand{\MinimizeWithinBucket}[0]{\mathop{\textsc{Minimize1}}}
\newcommand{\MinimizeAcrossBuckets}[0]{\mathop{\textsc{Minimize2}}}
\newcommand{\false}[0]{\mathord{\mathit{false}}}
\newcommand{\true}[0]{\mathord{\mathit{true}}}

\newcommand{\SubSubSection}[0]{\subsubsection}  % icde
\def\phi{\varphi}

\newcommand{\Set}[1]{\mathord{\lbrace{#1}\rbrace}}

\def\|{\mathbin{\mid}}

\newcommand{\B}[0]{\mathord{\mathcal{B}}}

\renewcommand{\L}[0]{\mathord{\mathcal{L}}}
\newcommand{\Lb}[1]{\mathord{\mathcal{L}_{\mathrm{basic}}^{#1}}}

\newcommand{\NP}[0]{\mathord{\mathrm{NP}}}
\newcommand{\SharpP}[0]{\mathord{\#\mathrm{P}}}
\newcommand{\from}[0]{\mathbin{\leftarrow}}

\title{Worst-Case Background Knowledge for Privacy-Preserving Data Publishing}

\author{%
David J. Martin \and Daniel Kifer \and Ashwin Machanavajjhala
\and Johannes Gehrke
\and Joseph Y. Halpern
\\
Cornell University \\
\{djm, dkifer, mvnak, johannes, halpern\}@cs.cornell.edu \\
}

\begin{document}

\maketitle
\thispagestyle{empty}  % icde style

\begin{abstract}
\input{abstract}
\end{abstract}

\input{introduction}

\input{framework}

\input{implication}

\input{experiments}

\input{relatedwork}
\input{conclusions}

\bibliographystyle{plain}
{
\small
\bibliography{main}
}

\end{document}

%% file: abstract.tex
Recent work has shown the necessity of considering an attacker's background knowledge when reasoning about privacy in data publishing.  
However, in practice, the data publisher does not know what background knowledge the attacker possesses.  
Thus, it is important to consider the worst-case.  
In this paper, we initiate a formal study of worst-case background knowledge.  
We propose a language that can express any background knowledge about the data.
We provide a polynomial time algorithm to measure the amount of disclosure of sensitive information in the worst case, given that the attacker has at most $k$ pieces of information in this language.
We also provide a method to efficiently sanitize the data so that the amount of disclosure in the worst case is less than a specified threshold.

%% file: introduction.tex
\Section{Introduction}\label{sec:introduction}

\newcommand{\Nodes}[3]{ & }
\newcommand{\Lines}[2]{}
\DrawLines{
\renewcommand{\Nodes}[3]{
\begin{pspicture}[.97](0,0)(0,0)\cnode(-0.1,0.1){0}{#1X#2}\end{pspicture} & 
\begin{pspicture}[.97](0,0)(0,0)\cnode(0.1,0.1){0}{#1S#3}\end{pspicture}
}
\renewcommand{\Lines}[2]{\begin{pspicture}[.97](0,0)(0,0)\ncline[linestyle=solid]{#1X#2}{#1S#2}\end{pspicture}}
}

\newcommand{\Hline}[0]{\cline{1-1}\cline{2-4}\cline{7-7}}
\newcommand{\Nodesa}[3]{\Nodes{#1}{#2}{#3}}
\newcommand{\Linesa}[2]{\Lines{#1}{#2}}

\begin{figure*}[t]
\centering
%\BlockComment{
\begin{minipage}{2.1in}
\footnotesize{
\centering
\begin{tabular}{|c|c|c|c|c|}
\multicolumn{1}{c}{}  \!\!& \multicolumn{3}{c}{non-sensitive} \!\!& \multicolumn{1}{c}{sensitive} \!\!\\
\hline
\!\!\!\!\textbf{Name} \!\!\!\!&\!\!\!\! \textbf{Zip} \!\!\!\!&\!\!\!\! \textbf{Age} \!\!\!\!&\!\!\!\! \textbf{Sex} \!\!\!\!&\!\!\!\! \textbf{Disease} \!\!\!\!\\
\hline
\!\!\!\!Bob      \!\!\!\!&\!\!\!\! 14850    \!\!\!\!&\!\!\!\! 23 \!\!\!\!&\!\!\!\! M    \!\!\!\!&\!\!\!\! Flu \!\!\!\!\\
\hline
\!\!\!\!Charlie  \!\!\!\!&\!\!\!\! 14850    \!\!\!\!&\!\!\!\! 24 \!\!\!\!&\!\!\!\! M    \!\!\!\!&\!\!\!\! Flu \!\!\!\!\\
\hline
\!\!\!\!Dave     \!\!\!\!&\!\!\!\! 14850    \!\!\!\!&\!\!\!\! 25 \!\!\!\!&\!\!\!\! M   \!\!\!\!&\!\!\!\! Lung Cancer \!\!\!\!\\
\hline
\!\!\!\!Ed       \!\!\!\!&\!\!\!\! 14850    \!\!\!\!&\!\!\!\! 27 \!\!\!\!&\!\!\!\! M   \!\!\!\!&\!\!\!\! Lung Cancer \!\!\!\!\\
\hline
\!\!\!\!Frank    \!\!\!\!&\!\!\!\! 14853    \!\!\!\!&\!\!\!\! 29 \!\!\!\!&\!\!\!\! M  \!\!\!\!&\!\!\!\! Mumps \!\!\!\!\\
\hline
\!\!\!\!Gloria   \!\!\!\!&\!\!\!\! 14850    \!\!\!\!&\!\!\!\! 21 \!\!\!\!&\!\!\!\! F  \!\!\!\!&\!\!\!\! Flu \!\!\!\!\\
\hline
\!\!\!\!Hannah   \!\!\!\!&\!\!\!\! 14850    \!\!\!\!&\!\!\!\! 22 \!\!\!\!&\!\!\!\! F  \!\!\!\!&\!\!\!\! Flu \!\!\!\!\\
\hline
\!\!\!\!Irma     \!\!\!\!&\!\!\!\! 14853    \!\!\!\!&\!\!\!\! 24 \!\!\!\!&\!\!\!\! F  \!\!\!\!&\!\!\!\! Breast Cancer \!\!\!\!\\
\hline
\!\!\!\!Jessica  \!\!\!\!&\!\!\!\! 14853    \!\!\!\!&\!\!\!\! 26 \!\!\!\!&\!\!\!\! F  \!\!\!\!&\!\!\!\! Ovarian Cancer \!\!\!\!\\
\hline
\!\!\!\!Karen    \!\!\!\!&\!\!\!\! 14853    \!\!\!\!&\!\!\!\! 28 \!\!\!\!&\!\!\!\! F  \!\!\!\!&\!\!\!\! Heart Disease \!\!\!\!\\
\hline
\end{tabular}
%\Linesa{A}{6}
%\Linesa{A}{7}
%\Linesa{A}{8}
%\Linesa{A}{9}
%\Linesa{A}{10}
%\Linesa{A}{11}
%\Linesa{A}{12}
%\Linesa{A}{13}
%\Linesa{A}{14}
%\Linesa{A}{15}
%\end{center}
\caption{\normalsize{Original table}}\label{fig:original}
}
\end{minipage}
%}
%\hspace{.03in} 
\ \ 
\begin{minipage}{2.1in}
\footnotesize{
\centering
\begin{tabular}{|c|c|c|c|c|}
\multicolumn{1}{c}{} \!\!& \multicolumn{3}{c}{non-sensitive} \!\!& \multicolumn{1}{c}{sensitive} \!\!\\
\hline
\!\!\!\!\textbf{Name} \!\!\!\!&\!\!\!\! \textbf{Zip} \!\!\!\!&\!\!\!\! \textbf{Age} \!\!\!\!&\!\!\!\! \textbf{Sex} \!\!\!\!&\!\!\!\! \textbf{Disease} \!\!\!\!\\
\hline
\!\!\!\!   \!\!\!\!&\!\!\!\!     \!\!\!\!&\!\!\!\!  \!\!\!\!&\!\!\!\!  \!\!\!\!&\!\!\!\! Flu \!\!\!\!\\
\!\!\!\!   \!\!\!\!&\!\!\!\!     \!\!\!\!&\!\!\!\!  \!\!\!\!&\!\!\!\!  \!\!\!\!&\!\!\!\! Lung Cancer \!\!\!\!\\
\!\!\!\! *  \!\!\!\!&\!\!\!\! 1485*    \!\!\!\!&\!\!\!\! 2* \!\!\!\!&\!\!\!\! M \!\!\!\!&\!\!\!\! Mumps \!\!\!\!\\
\!\!\!\!   \!\!\!\!&\!\!\!\!     \!\!\!\!&\!\!\!\!  \!\!\!\!&\!\!\!\!  \!\!\!\!&\!\!\!\! Flu \!\!\!\!\\
\!\!\!\!   \!\!\!\!&\!\!\!\!     \!\!\!\!&\!\!\!\!  \!\!\!\!&\!\!\!\!  \!\!\!\!&\!\!\!\! Lung Cancer \!\!\!\!\\
\hline
\!\!\!\!  \!\!\!\!&\!\!\!\!     \!\!\!\!&\!\!\!\!  \!\!\!\!&\!\!\!\!  \!\!\!\!&\!\!\!\! Flu \!\!\!\!\\
\!\!\!\!  \!\!\!\!&\!\!\!\!     \!\!\!\!&\!\!\!\!  \!\!\!\!&\!\!\!\!  \!\!\!\!&\!\!\!\! Breast Cancer \!\!\!\!\\
\!\!\!\! * \!\!\!\!&\!\!\!\! 1485*    \!\!\!\!&\!\!\!\! 2* \!\!\!\!&\!\!\!\! F \!\!\!\!&\!\!\!\! Flu \!\!\!\!\\
\!\!\!\!  \!\!\!\!&\!\!\!\!     \!\!\!\!&\!\!\!\!  \!\!\!\!&\!\!\!\!  \!\!\!\!&\!\!\!\! Heart Disease \!\!\!\!\\
\!\!\!\!  \!\!\!\!&\!\!\!\!     \!\!\!\!&\!\!\!\!  \!\!\!\!&\!\!\!\!  \!\!\!\!&\!\!\!\! Ovarian Cancer \!\!\!\!\\
\hline
\end{tabular}
\caption{
\normalsize{5-anonymous table}}\label{fig:5-anon}}
\end{minipage}
%\hspace{.03in} 
\ \
%\hfill
\begin{minipage}{2.4in}
\footnotesize{
\centering
\begin{tabular}{|c|c|c|c|cc|c|}
\multicolumn{1}{c}{}  \!\!& \multicolumn{3}{c}{non-sensitive} \!\!& \multicolumn{2}{c}{} \!\!& \multicolumn{1}{c}{sensitive} \!\!\\
\Hline
\!\!\!\!\textbf{Name} \!\!\!\!&\!\!\!\! \textbf{Zip} \!\!\!\!&\!\!\!\! \textbf{Age} \!\!\!\!&\!\!\!\! \textbf{Sex} \!\!\!\!&\!\!\!\! \!\!\!\!&\!\!\!\!  \!\!\!\!&\!\!\!\! \textbf{Disease} \!\!\!\!\\
\Hline
\!\!\!\!Bob      \!\!\!\!&\!\!\!\! 14850    \!\!\!\!&\!\!\!\! 23 \!\!\!\!&\!\!\!\! M \!\!\!\!&\!\!\!\! \Nodes{B}{6}{7}   \!\!\!\!&\!\!\!\! Flu \!\!\!\!\\
\!\!\!\!Charlie  \!\!\!\!&\!\!\!\! 14850    \!\!\!\!&\!\!\!\! 24 \!\!\!\!&\!\!\!\! M \!\!\!\!&\!\!\!\! \Nodes{B}{7}{9}   \!\!\!\!&\!\!\!\! Lung Cancer \!\!\!\!\\
\!\!\!\!Dave     \!\!\!\!&\!\!\!\! 14850    \!\!\!\!&\!\!\!\! 25 \!\!\!\!&\!\!\!\! M \!\!\!\!&\!\!\!\! \Nodes{B}{8}{10}  \!\!\!\!&\!\!\!\! Mumps \!\!\!\!\\
\!\!\!\!Ed       \!\!\!\!&\!\!\!\! 14850    \!\!\!\!&\!\!\!\! 27 \!\!\!\!&\!\!\!\! M \!\!\!\!&\!\!\!\! \Nodes{B}{9}{6}   \!\!\!\!&\!\!\!\! Flu \!\!\!\!\\
\!\!\!\!Frank    \!\!\!\!&\!\!\!\! 14853    \!\!\!\!&\!\!\!\! 29 \!\!\!\!&\!\!\!\! M \!\!\!\!&\!\!\!\! \Nodes{B}{10}{8}  \!\!\!\!&\!\!\!\! Lung Cancer \!\!\!\!\\
\Hline
\!\!\!\!Gloria   \!\!\!\!&\!\!\!\! 14850    \!\!\!\!&\!\!\!\! 21 \!\!\!\!&\!\!\!\! F \!\!\!\!&\!\!\!\! \Nodes{B}{11}{12} \!\!\!\!&\!\!\!\! Flu \!\!\!\!\\
\!\!\!\!Hannah   \!\!\!\!&\!\!\!\! 14850    \!\!\!\!&\!\!\!\! 22 \!\!\!\!&\!\!\!\! F \!\!\!\!&\!\!\!\! \Nodes{B}{12}{13} \!\!\!\!&\!\!\!\! Breast Cancer \!\!\!\!\\
\!\!\!\!Irma     \!\!\!\!&\!\!\!\! 14853    \!\!\!\!&\!\!\!\! 24 \!\!\!\!&\!\!\!\! F \!\!\!\!&\!\!\!\! \Nodes{B}{13}{11} \!\!\!\!&\!\!\!\! Flu \!\!\!\!\\
\!\!\!\!Jessica  \!\!\!\!&\!\!\!\! 14853    \!\!\!\!&\!\!\!\! 26 \!\!\!\!&\!\!\!\! F \!\!\!\!&\!\!\!\! \Nodes{B}{14}{15} \!\!\!\!&\!\!\!\! Heart Disease \!\!\!\!\\
\!\!\!\!Karen    \!\!\!\!&\!\!\!\! 14853    \!\!\!\!&\!\!\!\! 28 \!\!\!\!&\!\!\!\! F \!\!\!\!&\!\!\!\! \Nodes{B}{15}{14} \!\!\!\!&\!\!\!\! Ovarian Cancer \!\!\!\!\\
\Hline
\end{tabular}
\Lines{B}{6}
\Lines{B}{7}
\Lines{B}{8}
\Lines{B}{9}
\Lines{B}{10}
\Lines{B}{11}
\Lines{B}{12}
\Lines{B}{13}
\Lines{B}{14}
\Lines{B}{15}
\caption{\normalsize{Bucketized table}}\label{fig:anonymized}
%\end{center}
}
\end{minipage}
\end{figure*}

We consider the following situation.
A data publisher (such as a hospital) has collected useful information about a group of individuals (such as patient records that would help medical researchers) and would like to publish this data while preserving the privacy of the individuals involved.
The information is stored as a table (as in Figure \ref{fig:original}) where each record corresponds to a unique individual and contains a sensitive attribute (e.g., disease) and some non-sensitive attributes (e.g., address, gender, age) that might be learned using externally available data (e.g., phone books, birth records).
The data publisher would like to limit the disclosure of the sensitive values of the individuals in order to defend against an attacker who possibly already knows some facts about the table.
Our goal in this paper is to quantify the precise effect of background knowledge possessed by an attacker on the amount of disclosure and to provide algorithms to check and ensure that the amount of disclosure is less than a specified threshold.

The problem we solve is of real and practical importance; 
an egregious example of a privacy breach was the discovery of the medical
records of the Governor of Massachusetts from an easily accessible and
supposedly anonymized dataset.  All that was needed was to link it to
voter registration records \cite{sweeney02:kAnon}.
To defend against such attacks, Samarati and Sweeney \cite{samaratiS98:protecting} introduced a privacy criterion called $k$-anonymity which requires that each individual be indistinguishable (with respect to the non-sensitive attributes) from at least $k - 1$ others.
This is done by grouping individuals into \emph{buckets} of size at least $k$, and then permuting the sensitive values in each bucket and sufficiently masking their externally observable non-sensitive attributes.  Figure \ref{fig:5-anon} depicts a table that is a $5$-anonymous version of the table in Figure \ref{fig:original}.
Figure \ref{fig:anonymized} depicts the permutation of sensitive values that was used to construct this table.

However, $k$-anonymity does not adequately protect the privacy of an individual;%
\footnote{Indeed, the definition of $k$-anonymity does not even mention the sensitive attribute!}
for example, when all individuals in a bucket have the same disease, the disease of the individuals in that bucket is disclosed regardless of the bucket size.
Even when there are multiple diseases in the same bucket, the frequencies of the diseases in the bucket still matter when an attacker has some background knowledge about the particular individuals in the table.
Suppose the data publisher has published the $5$-anonymous table as depicted in Figure \ref{fig:5-anon}.
Consider an attacker Alice who would like to learn the diseases of all her friends and neighbors. 
One of her neighbors is Ed, a 27 year-old male living in Ithaca (zip
code 14850).  Alice knows that Ed is in the hospital that published the
anonymized dataset in Figure \ref{fig:anonymized}, and she wants to find out
Ed's disease. 
Using her knowledge of Ed's age, gender, and zip-code, Alice can identify the bucket in the anonymized table that Ed belongs to (namely, the first bucket).
Alice does not know which disease listed within that bucket is Ed's since the sensitive values were permuted.
Therefore, without additional knowledge, Alice's estimate of the probability that Ed has lung cancer is $2/5$. 
But suppose Alice knows that Ed had mumps as a child and is therefore extremely unlikely to get it again. 
After ruling out this possibility, the probability that Ed has lung cancer increases to $1/2$. 
Now, if Alice also somehow discovers that Ed does not have flu, then the fact that he has lung cancer becomes certain. 
Here, two pieces of knowledge of the form ``Ed does not have X'' were
enough to fully disclose Ed's disease. 
To guard against this, Machanavajjhala et al. 
\cite{machanavajjhalaGKV06:l-diversity} proposed a privacy criterion called $\ell$-diversity that ensures that it takes at least $\ell-1$ such
pieces of information to sufficiently disclose the sensitive value of any individual. 
The main idea is to require that, for each bucket, the $\ell$ most frequent sensitive values are roughly equi-probable.

$\ell$-diversity focuses on one type of background knowledge: knowledge of the form
  ``individual X does not have sensitive value Y''.  
But an attacker might well have other types of background knowledge.  
For example, 
suppose Alice lives across the street from a married couple,
Charlie and Hannah, who were both taken to the hospital. 
Once again, using her knowledge of their genders, ages and zip-codes, Alice can identify the buckets Charlie and Hannah belong to.
Without additional background knowledge, Alice thinks that Charlie has the flu with
probability $2/5$. 
But suppose that Alice knows that Hannah has had a flu shot recently but Charlie has not.  
Believing Hannah's immunity to the flu to be much stronger than Charlie's and knowing that they live together, Alice deduces that if Hannah has succumbed to the flu then it is extremely likely that Charlie has as well. 
This knowledge allows her to update her probability that Charlie has
the flu to $10/19$. We show how these probabilities are computed in Section
$\ref{sec:implication}$. 
$\ell$-diversity does not guard against the type of background knowledge in this example.

It is thus clear that we need a more general-purpose framework that can capture knowledge of \emph{any} property of the underlying table that an attacker might know.
Moreover, unlike in the two examples above where we knew Alice's background knowledge, we will not assume that we know exactly what the attacker knows.  
We therefore take the following approach.
In Section \ref{sec:framework}, 
we propose a language that is expressive enough to capture any property of the sensitive values in a table.
This language enables us to decompose background knowledge into basic units of information.
Then, given an anonymized version of the table, we can quantify the worst-case disclosure risk posed by an attacker with $k$ such units of information;  $k$ can be thought of as a bound on the power of an attacker.
In Section \ref{sec:implication}, we show how to efficiently preserve privacy by ensuring that the \emph{worst-case} (i.e., maximum) disclosure for \emph{any} $k$ pieces of information is less than a specified threshold.
Furthermore, we show to integrate our techniques into existing frameworks to find a ``minimally sanitized'' table for which the maximum disclosure is less than a specified threshold.
We present experiments in Section \ref{sec:experiments}, related work in Section \ref{sec:relatedwork}, and
we conclude in Section \ref{sec:conclusions}.

To the best of our knowledge, this is the first such formal analysis of the effect of unknown background knowledge on the disclosure of sensitive information.

%% file: framework.tex
\Section{Framework}\label{sec:framework}
We begin by modeling the data publishing situation formally.  Let $P$ be a (finite) set of
people.  For each $p \in P$, we associate a tuple
$t_p$ which has one sensitive attribute $S$ (e.g., disease) with finite domain and one or more non-sensitive attributes.  
We overload notation and use $S$ to represent both the sensitive attribute and its domain.
The data publisher has a table $T$, which is a set of tuples corresponding to a subset of $P$.  
The publisher would like publish $T$ in a form that protects the sensitive information of any individual from an attacker with background knowledge that can be expressed in a language $\L$.
(We propose such a language to express background knowledge in Section $\ref{sec:language}$.)

\SubSection{Bucketization}
We first need to carefully describe how the published data is constructed from the underlying table
if we are to correctly interpret this published data.
That is, we need to specify a sanitization method.  
We briefly describe two popular sanitization methods.
\begin{itemize}
\item
The first, which we term \emph{bucketization} $\cite{xiao06:anatomy}$, is to partition the tuples in $T$ into \emph{buckets}, and then to separate the sensitive attribute from the non-sensitive ones by randomly permuting the sensitive attribute values within each bucket.  The sanitized data then consists of the buckets with permuted sensitive values.
\item
The second sanitization technique is \emph{full-domain generalization} $\cite{sweeney02:kAnon}$, where we coarsen the non-sensitive attribute domains.  The sanitized data consists of the coarsened table along with generalization used.
Note that, unlike bucketization, the exact values of the non-sensitive attributes are not released; only the coarsened values are released.
\end{itemize}
Note that if the attacker knows the set of people in the table and their non-sensitive values, then full-domain generalization and bucketization are equivalent.
In this paper, we use bucketization as the method of constructing the published data from the original table $T$, although all our results hold for full-domain generalization as well.
We plan to extend our algorithms to work for other sanitization techniques, such as data swapping \cite{daleniusR82:swapping} (which, like bucketization, also permutes the sensitive values, but in more complex ways) and suppression \cite{samaratiS98:protecting}, in the future.

We now specify our notion of bucketization more formally.
Given a table $T$, we partition the tuples into buckets (i.e., horizontally partition the table $T$ according
to some scheme), and within each bucket, we apply an independent random permutation to the column containing $S$-values.  
The resulting set of buckets, denoted by $\B$, is then published.
For example, if the underlying table $T$ is as depicted in Figure $\ref{fig:original}$, then the publisher might publish bucketization $\B$ as depicted in Figure $\ref{fig:anonymized}$.  Of course, for added privacy, the publisher can completely mask the identifying attribute (Name) and may partially mask some of the other non-sensitive attributes (Age, Sex, Zip).

For a bucket $b \in \B$, we use the following notation. 
\begin{figure}[h]
\begin{minipage}{3.2in}
\centering
\begin{tabular}{|cl|}
\hline
$P_b$ & set of people $p \in P$ with tuples $t_p \in b$\\
$n_b$ & number of tuples in $b$\\
$n_b(s)$ & frequency of sensitive value $s \in S$ in $b$\\
$s_b^0, s_b^1, \dots$ & sensitive values in decreasing order \\
                      & of frequency in $b$\\
\hline
\end{tabular}
\end{minipage}
\end{figure}

\input{language}

\SubSection{Disclosure}

Having specified how the bucketization $\B$ is constructed from the underlying table $T$ and how an attacker's knowledge about sensitive information can be expressed in language $\Lb{k}$, we are now in a position to define our notion of disclosure precisely.

\begin{definition}[Disclosure risk]
The \emph{disclosure risk} of bucketization $\B$ with respect to background knowledge represented by some formula $\phi$ in language $\Lb{k}$ is
$$
\max_{t_p \in T, s \in S} \Pr(t_p[S] = s \| \B \land \phi)
$$
That is, \textit{disclosure risk} is the likelihood of the most highly predicted sensitive attribute assignment.
\end{definition}
\begin{definition}[Maximum disclosure]
The \emph{maximum disclosure} of bucketization $\B$ with respect to language $\Lb{k}$ that expresses background knowledge is
$$
\max_{t_p \in T, s \in S, \phi \in \Lb{k}} \Pr(t_p[S] = s \| \B \land \phi)
$$
\end{definition}
By our assumptions in \ref{sec:language}, we compute $\Pr(t_p[S] = s \| \B \land \phi)$ by considering the set of all tables consistent with bucketization $\B$ and with background knowledge $\phi$ and then taking the fraction of those tables that satisfy $t_p[S] = s$.  
Using this, the maximum disclosure of the bucketization in Figure \ref{fig:anonymized} with respect to $\Lb{1}$ turns out to be $\frac{10}{19}$, and occurs when $\phi$ is $t_{p'} = s' \to t_{p} = s$ where $p$ is a person in the first bucket, $p'$ is a person in the second bucket, and $s$ and $s'$ are both flu.
Our goal is to develop general techniques to:
\begin{enumerate}
\item
efficiently calculate the maximum disclosure for any given bucketization, and 
\item
efficiently find a ``minimally sanitized'' bucketization\footnote{We will make precise the notion of ``minimally sanitized'' in Section $\ref{sec:imp:monotonicity}$; we want ``minimal sanitization'' in order to preserve the utility of the data.} (or the set of all minimally sanitized bucketizations) for which the maximum disclosure is below a specified threshold (if any exist).
\end{enumerate}

%% file: language.tex
\SubSection{Background Knowledge}\label{sec:language}

We pessimistically assume that the attacker has managed to obtain complete information about 
which individuals have records in the table, what their non-sensitive data is, and which buckets in the bucketization these records fall into.
That is, we assume that the attacker knows $P_b$, the set of people in bucket $b$, for each $b \in \B$, and knows $t_p[X]$ for every person $p$ in the table and every non-sensitive attribute $X$.
We call this \emph{full identification information}.
One way of obtaining identification information in practice is to link quasi-identifying non-sensitive attributes published in the bucketization (e.g., address, gender, age) with publicly available data (e.g., phone directories, birth records) \cite{sweeney02:kAnon}.

We make the standard random worlds assumption \cite{bacchusGHK96:randomworld}: in the absence of any further knowledge, we consider all tables consistent with this bucketization to be equally likely.
That is, the probability of $t_p \in b$ having $s$ for its sensitive attribute is $n_b(s)/n_b$ since each assignment of sensitive attributes to tuples within a bucket is equally likely.

We now need to consider knowledge beyond the identification information that an attacker might possess.  
We assume that this further knowledge is the knowledge that the underlying table satisfies a given \emph{predicate} on tables.
That is, the attacker knows that the underlying table is among the set of tables satisfying the given predicate.
This is a rather general assumption.  For example, ``the average age of heart disease patients in the table is 48 years'' could be one such predicate.
In order to quantify the power of such knowledge, we use the notion of a \emph{basic unit} of knowledge, and
we propose a language which consists of finite conjunctions of such basic units.
Given full identification information, we desire that any predicate on tables be expressible using a conjunction of the basic units that we propose.
We employ a very simple propositional syntax.
\begin{definition}[Atoms]
An \emph{atom} is a formula of the form $t_p[S] = s$, for some value $s \in S$ and person $p \in P$ with tuple $t_p \in T$.
We say that atom $t_p[S] = s$ \emph{involves} person $p$ and value $s$.
\end{definition}
The interpretation of atoms is obvious:
$t_{\mathrm{Jack}}[\textrm{Disease}] = \mathrm{flu}$ says that the Jack's tuple has the value $\mathrm{flu}$ for the sensitive attribute $\textrm{Disease}$.

The basic units of knowledge in our language are \emph{basic implications}, defined below.
\begin{definition}[Basic implications]
A \emph{basic implication} is a formula of the form 
$$(\land_{i \in [m]} A_i) \to (\lor_{j \in [n]} B_j)$$
for some $m \geq 1, n \geq 1$ and atoms $A_i, B_j$, $i \in [m], j \in [n]$
(note that we use the standard notation $[n]$ to denote the set $\Set{0, \dots, n - 1}$).
\end{definition}
The fact that basic implications are a sufficiently expressive ``basic unit'' of knowledge is made precise by the following theorem.%
\footnote{\IfTechReport{See appendix for proofs.}{See \cite{martinKMGH06:worstCaseTR} for proofs.}}
\begin{theorem}[Completeness]\label{thm:basic-impl}
Given full identification information and any predicate on tables, one can express the knowledge that the underlying table satisfies the identification information and the given predicate using a finite conjunction of basic implications.
\end{theorem}
Hence we can model arbitrarily powerful attackers.%
\footnote{%
A major shortcoming of the $\ell$-diversity definition was that its choice of ``basic unit'' of knowledge was essentially negated atoms (i.e., $\lnot t_p[S] = s$) which cannot capture all properties of the underlying table.  For example, negations cannot express basic implications in general.
}
Consider an attacker who knows the disease of every person in the table except for Bob.
Then publishing any bucketization will reveal Bob's disease.
To avoid pathological and unrealistic cases like this, we need to assume a bound on the power of an attacker.
We model attackers with bounded power by limiting the number of basic implications that the attacker knows. 
That is, the attacker knows a single formula from language $\Lb{k}$ defined below.
\begin{definition}
$\Lb{k}$ is the language consisting of conjunctions of $k$ basic implications.  
That is, $\Lb{k}$ consists of formulas of the form $\land_{i \in [k]} \phi_i$ where each $\phi_i$ is a basic implication.
\end{definition}
$k$ can thus be viewed as a bound on the attacker's power and can be increased to provide more conservative privacy guarantees.

Note that our choice of basic implications for the ``basic unit'' of our language has important consequences on our assumptions about the attacker's power.  
In particular, some properties of the underlying table might require a large number of basic implications to express.  
Since basic implications are essentially CNF clauses with at least one negative atom, our language suffers from an exponential blowup in the number of basic units required to express arbitrary DNF formulas. 
It may be that other choices of basic units may lead to equally expressive
languages while at the same time requiring fewer basic units to express
certain natural properties, and we consider this an important direction for future research.
Nevertheless, many natural types of background knowledge have succinct representations using basic implications.
For example, 
Alice's knowledge that ``if Hannah has the flu, then Charlie also has the flu'' is simply the basic implication
$$t_{\mathrm{Hannah}}[\mathrm{Disease}] = \mathrm{flu} \to t_{\mathrm{Charlie}}[\mathrm{Disease}] = \mathrm{flu}$$ 
And the knowledge that ``Ed does not have flu'' is
$$t_{\mathrm{Ed}}[\mathrm{Disease}] = \mathrm{flu} \to t_{\mathrm{Ed}}[\mathrm{Disease}] = \mathrm{ovarian~cancer}$$ 
In general, we can represent $\lnot t[S] = s$ by $(t[S] = s) \to (t[S] = s')$ for any choice of $s' \neq s$ since each tuple has exactly one sensitive attribute value.  

Note that maintaining privacy when there is dependence between sensitive values, especially \emph{across buckets}, is a problem that has not been previously addressed in the privacy literature.
The assignments of individuals to sensitive values in different buckets are not necessarily independent.  
As we saw in the example with Hannah and Charlie, fixing a particular assignment in one bucket could affect what assignments are possible in another.  
One of the contributions of this paper is that we provide a polynomial time algorithm for computing the maximum disclosure even when the attacker has knowledge of such dependencies.

%% file: implication.tex
\Section{Checking And Enforcing Privacy}\label{sec:implication}

In Section $\ref{sec:language}$, we defined basic implications as the ``unit of knowledge'' and showed that this was a fully expressive (in the presence of full identification information) and reasonable choice.
We now show how to efficiently calculate and limit maximum disclosure against an attacker who has full identification information and has up to $k$ additional pieces of background knowledge (i.e., up to $k$ basic implications).
In order to do this, we will show in Theorem \ref{thm:specialform} that there is a set of $k$ basic implications that maximizes disclosure with respect to $\Lb{k}$.  Furthermore, each such implication has \emph{only one atom in the antecedent and one atom in the consequent}.
This motivates the following definition.
\begin{definition}[Simple implications]
A \emph{simple implication} is a formula of the form $A \to B$ for some atoms $A, B$.
\end{definition}

\SubSection{Hardness of computing disclosure risk}
Unfortunately, naive methods for computing the maximum disclosure will not work -- in fact, we can show that computing the disclosure risk of a given bucketization with respect to a given set of $k$ simple implications is $\SharpP$-hard.
Note that $k$ simple implications can be written in $2$-CNF, for which satisfiability is easily checkable.  Complexity is introduced in trying to \emph{simultaneously} satisfy the $k$ implications \emph{and} the given bucketization.
In fact, deciding whether a given bucketization is consistent with a set of $k$ simple implications is $\NP$-complete.
\begin{theorem}\label{thm:impnp}
Given as input bucketization $\B$ %(represented as $P_b$ and $n_b(s)$ for each $b \in \B$ and each $s \in S$)
and a conjunction of simple implications $\phi$,
the problem of deciding if $\B$ and $\phi$ are both satisfiable by some table $T$ is $\NP$-complete.
Moreover, given an atom $C$ as further input,
the problem of computing $\Pr(C \| \B \land \land_{i \in [k]} \phi_i)$ is $\SharpP$-complete.
\end{theorem}

\SubSection{A special form for maximum disclosure}
It turns out that, despite the hardness results above, computing the \emph{maximum} disclosure with respect to language $\Lb{k}$ can be done in polynomial time.
The key insight is summarized in Theorem \ref{thm:specialform}.
\begin{theorem}\label{thm:specialform}
For any bucketization, there is a set of \emph{$k$ simple implications, all sharing the same consequent,} such that the conjunction of these $k$ simple implications maximizes disclosure with respect to $\Lb{k}$.
\end{theorem}
This insight is tremendously useful in devising a polynomial-time dynamic programming algorithm for computing the maximum disclosure with respect to $\Lb{k}$ as it allows us to restrict our attention to sets of $k$ simple implications of the form $(t_{p_i}[S] = s_i) \to (t_p[S] = s)$ for people $p, p_i \in P$, and values $s, s_i \in S$, $i \in [k]$.
The proof of Theorem \ref{thm:specialform} follows from the following two lemmas.

\begin{lemma}\label{thm:impmaster}
For any formulas $\psi, \phi, \theta_i, \phi_i$,
$$
\begin{array}{lcl}
\multicolumn{3}{l}
{\Pr(\phi \| \psi \land (\land_{i \in [k]} (\theta_i \to \phi_i)))}
\\
& \leq & \Pr(\phi \| \psi \land (\land_{i \in [k]} (\theta_i \to \phi))) \\
\end{array}
$$
\end{lemma}

Starting with any set of $k$ basic implications that maximize disclosure,%
\footnote{There always exists some set of $k$ basic implications that maximize disclosure since there are only finitely many atoms and therefore $\Lb{k}$ is finite.} 
Lemma $\ref{thm:impmaster}$ enables us to replace the consequent in all the basic implications by a single common atom (namely the atom corresponding to the highest predicted assignment of sensitive value to an individual), while still maintaining maximum disclosure.

\begin{lemma}\label{thm:basiclhs}
For any formulas $\psi, B, \theta_i$, where $B$ is an atom and $\theta_i$ is a conjunction of atoms, there
exist atoms $A_i$ such that 
$$
\begin{array}{lcl}
\multicolumn{3}{l}
{\Pr(B \| \psi \land (\land_{i \in [k]} (\theta_i \to B)))}
\\
& \leq & \Pr(B \| \psi \land (\land_{i \in [k]} (A_i \to B))). \\
\end{array}
$$
\end{lemma}

Next, Lemma $\ref{thm:basiclhs}$ allows us to replace the antecedent of each of the resulting implications by an atom (possibly with a different atom for each implication), while still maintaining maximum disclosure.

In both Lemmas $\ref{thm:impmaster}$ and $\ref{thm:basiclhs}$, we use $\psi$ to represent the attacker's knowledge about the bucketization $\B$.  
However, it is worthwhile pointing out that neither lemma places any restriction on $\psi$ or on the underlying probability distribution.
This makes the results presented here extremely general and powerful because \emph{they characterize the form of background knowledge that maximizes disclosure risk for any form of anonymization and for any additional background knowledge}.

The main idea behind the proof of Lemma \ref{thm:impmaster} (and also Lemma \ref{thm:basiclhs}) can be illustrated as follows.
Consider a bucketization $\B$.
Let $(t_{p_i}[S] = s_i) \to (t_{p'_i}[S] = s'_i)$, for $i \in \Set{0, 1}$, be two simple implications which maximize the disclosure of $\B$ with respect to $\Lb{2}$.
For convenience, we let $A_i$ denote the atom $t_{p_i}[S] = s_i$ and $B_i$ the atom $t_{p'_i}[S] = s'_i$.
Let $C$ be the atom $t_p[S] = s$ such that $\Pr(C \| \B \land (\land_{i \in [2]} (A_i \to B_i)))$ is the maximum disclosure.

Now let us restrict our attention to the set of tables consistent with $\B$.
Let $\mathcal{T}_1$ be the set of tables satisfying the simple implications $A_0 \to B_0$ and $A_1 \to B_1$, and let $\mathcal{T}_2$ be the set of tables satisfying $A_0 \to C$ and $A_1 \to C$.
Figure \ref{fig:tt} is a diagrammatic representation of $\mathcal{T}_1$ and $\mathcal{T}_2$.
Each row in the the truth table on the left (resp., right) in Figure \ref{fig:tt} represents a subset of $\mathcal{T}_1$ (resp., $\mathcal{T}_2$).
The variables $a, b, c, d, e, f, g, h$ in the left-most (resp., $a, b, d', f', h'$ in the right-most) column represents the size of the corresponding set.
For example, the set of tables represented by the second row is the set of tables that satisfy the atom $C$ but do not satisfy $A_0$ and $A_1$, and the number of such of tables is $b$.

\begin{figure}[t]
\centering
\footnotesize{
\begin{tabular}{c|ccccc||c||ccccc|c}
  \!\!\!& \multicolumn{5}{c||}{$\land_{i \in [2]} (A_i \to B_i)$} & 
  \!\!\!& \multicolumn{5}{c|}{$\land_{i \in [2]} (A_i \to C)$} \\
\hline\hline
 \!\!\!&\!\!\! $A_0$ \!\!\!&\!\!\! $A_1$ \!\!\!&\!\!\! $B_0$ \!\!\!&\!\!\! $B_1$ \!\!\!&\!\!\! $C$ \!\!\!&\!\!\! \!\!\!&\!\!\! $A_0$ \!\!\!&\!\!\! $A_1$ \!\!\!&\!\!\! $B_0$ \!\!\!&\!\!\! $B_1$ \!\!\!&\!\!\! $C$ \!\!\!&\!\!\! \\
\hline\hline
$a$ \!\!\!&\!\!\! 0 \!\!\!&\!\!\! 0 \!\!\!&\!\!\! * \!\!\!&\!\!\! * \!\!\!&\!\!\! 0 \!\!\!&\!\!\! $=$         \!\!\!&\!\!\! 0 \!\!\!&\!\!\! 0 \!\!\!&\!\!\! * \!\!\!&\!\!\! * \!\!\!&\!\!\! 0 \!\!\!&\!\!\! $a$ \\\hline
$b$ \!\!\!&\!\!\! 0 \!\!\!&\!\!\! 0 \!\!\!&\!\!\! * \!\!\!&\!\!\! * \!\!\!&\!\!\! 1 \!\!\!&\!\!\! $=$         \!\!\!&\!\!\! 0 \!\!\!&\!\!\! 0 \!\!\!&\!\!\! * \!\!\!&\!\!\! * \!\!\!&\!\!\! 1 \!\!\!&\!\!\! $b$ \\\hline
$c$ \!\!\!&\!\!\! 0 \!\!\!&\!\!\! 1 \!\!\!&\!\!\! * \!\!\!&\!\!\! 1 \!\!\!&\!\!\! 0 \!\!\!&\!\!\!           \!\!\!&\!\!\!   \!\!\!&\!\!\!   \!\!\!&\!\!\!   \!\!\!&\!\!\!   \!\!\!&\!\!\!   \!\!\!&\!\!\! \\\hline
$d$ \!\!\!&\!\!\! 0 \!\!\!&\!\!\! 1 \!\!\!&\!\!\! * \!\!\!&\!\!\! 1 \!\!\!&\!\!\! 1 \!\!\!&\!\!\! $\subseteq$ \!\!\!&\!\!\! 0 \!\!\!&\!\!\! 1 \!\!\!&\!\!\! * \!\!\!&\!\!\! * \!\!\!&\!\!\! 1 \!\!\!&\!\!\! $d'$ \\\hline
$e$ \!\!\!&\!\!\! 1 \!\!\!&\!\!\! 0 \!\!\!&\!\!\! 1 \!\!\!&\!\!\! * \!\!\!&\!\!\! 0 \!\!\!&\!\!\!           \!\!\!&\!\!\!   \!\!\!&\!\!\!   \!\!\!&\!\!\!   \!\!\!&\!\!\!   \!\!\!&\!\!\!   \!\!\!&\!\!\! \\\hline
$f$ \!\!\!&\!\!\! 1 \!\!\!&\!\!\! 0 \!\!\!&\!\!\! 1 \!\!\!&\!\!\! * \!\!\!&\!\!\! 1 \!\!\!&\!\!\! $\subseteq$ \!\!\!&\!\!\! 1 \!\!\!&\!\!\! 0 \!\!\!&\!\!\! * \!\!\!&\!\!\! * \!\!\!&\!\!\! 1 \!\!\!&\!\!\! $f'$ \\\hline
$g$ \!\!\!&\!\!\! 1 \!\!\!&\!\!\! 1 \!\!\!&\!\!\! 1 \!\!\!&\!\!\! 1 \!\!\!&\!\!\! 0 \!\!\!&\!\!\!           \!\!\!&\!\!\!   \!\!\!&\!\!\!   \!\!\!&\!\!\!   \!\!\!&\!\!\!   \!\!\!&\!\!\!   \!\!\!&\!\!\! \\\hline
$h$ \!\!\!&\!\!\! 1 \!\!\!&\!\!\! 1 \!\!\!&\!\!\! 1 \!\!\!&\!\!\! 1 \!\!\!&\!\!\! 1 \!\!\!&\!\!\! $\subseteq$ \!\!\!&\!\!\! 1 \!\!\!&\!\!\! 1 \!\!\!&\!\!\! * \!\!\!&\!\!\! * \!\!\!&\!\!\! 1 \!\!\!&\!\!\! $h'$ \\\hline
\end{tabular}
}\vspace{-0.5em}
\caption{Truth tables}\label{fig:tt}
\end{figure}

It is now clear from Figure \ref{fig:tt} that the implications $A_0 \to C$ and $A_1 \to C$ also produce the maximum disclosure as follows.
$\Pr(C \| \land_{i \in [2]} A_i \to B_i)
= \frac{b + d + f + h}{a + b + c + d + e + f + g + h}$
and
$\Pr(C \| \land_{i \in [2]} A_i \to C)
= \frac{b + d' + f' + h'}{a + b + d' + f' + h'}$. 
Also
$\frac{b + d + f + h}{a + b + c + d + e + f + h}
\leq \frac{b + d + f + h}{a + b + d + f + h} 
\leq \frac{b + d' + f' + h'}{a + b + d' + f' + h'}$
since $d \leq d'$, $f \leq f'$, and $h \leq h'$.
Thus $\Pr(C \| \land_{i \in [2]} A_i \to B_i) \leq \Pr(C \| \land_{i \in [2]} A_i \to C)$.

\SubSection{Computing maximum disclosure efficiently}\label{sec:imp:algorithm}

Having reduced our search space from sets of basic implications that could lead to maximum disclosure
to sets of simple implications with the same consequent, we are now in a position to create an efficient algorithm to compute the maximum disclosure.
We want to \emph{maximize} $\Pr(A \| {\B} \land \land_{i \in [k]} (A_i \to A))$ over all atoms $A, A_i$, $i \in [k]$.
Notice that for any atoms $A, A_i$, $i \in [k]$ such that $A$ and $\land_{i \in [k]} A_i \to A$ are consistent with bucketization $\B$ we have:
{
    \small
\begin{eqnarray*}
&&
\Pr(A \| \B \land (\land_{i \in [k]} A_i \to A)) \\
&& \hspace{.2cm} = \frac{\Pr(A \land (\land_{i \in [k]} (A_i \to A)) \| \B)}
         {\Pr((\land_{i \in [k]} (A_i \to A)) \| \B)}  \\
&& \hspace{.2cm} = \frac{\Pr(A \| \B)}
         {\Pr((\lnot{A} \land (\land_{i \in [k]} \lnot{A_i})) \lor A \| \B)}  \\
&& \hspace{.2cm} = \frac{\Pr(A \| \B)}
         {\Pr(\lnot{A} \land (\land_{i \in [k]} \lnot{A_i}) \| \B) + \Pr(A \| \B)}  \\
\end{eqnarray*}
}
So it suffices to construct an efficient algorithm to \emph{minimize}, over all atoms $A, A_i$, $i \in [k]$,
\begin{eqnarray}
&\frac{\Pr(\lnot{A} \land (\land_{i \in [k]} \lnot{A_i}) \| {\B})}{\Pr(A \| {\B})}.&
\label{eqn:tomin}
\end{eqnarray}

In Section \ref{sec:single-bucket-min}, we show how to minimize $\Pr(\land_{i \in [k]} \lnot A_i \| {\B})$ over atoms $A_i$ involving individuals in the same bucket. 
We use this in Section \ref{sec:alg:2} to provide a dynamic programming algorithm $\MinimizeWithinBucket$ that minimizes Formula (\ref{eqn:tomin}) over atoms $A, A_i$, $i \in [k]$ involving individuals in the same bucket.
Finally, in Section \ref{sec:alg:3}, we use $\MinimizeWithinBucket$ to construct another dynamic programming algorithm $\MinimizeAcrossBuckets$ to minimize Formula (\ref{eqn:tomin}) jointly over the entire bucketization.

\SubSubSection{Minimizing $\Pr(\land_{i \in [k]} \lnot A_i \| {\B})$ for one bucket}
\label{sec:single-bucket-min}

Consider all sets of $k$ atoms involving people whose tuples are in a single $b \in \B$.
Each set of $k$ atoms is associated with a tuple $(l, k_0, \dots, k_{l - 1})$, where $l$ is the number of people involved in the $k$ atoms, and $k_i$ is the number of atoms involving the $i$-th person.
We label the $k$ atoms $A_{i, j}$ for $i \in [l]$ and $j \in [k_i]$ such that atom $A_{i, j}$ is the $j$-th atom (out of $k_i$ atoms) involving the $i$-th person.
Lemma \ref{lem:impwithin} provides a closed form for the minimum value of $\Pr(\land_{i \in [k]} \lnot A_i \| {\B})$ over all sets of $k$ atoms associated with a particular $(l, k_0, \dots, k_{l - 1})$.

\begin{lemma}\label{lem:impwithin}
Let $b \in \B$ be any bucket.
Let $k$, $l$, and $k_0, k_1$, $\dots, k_{l - 1}$ be such that $k = \Sigma_{i \in [l]} k_i$ and $k_i \geq k_{i + 1}$ for all $i \in [l - 1]$.
Let $s_b^0, s_b^1, s_b^2, \dots$ be the sensitive values arranged in descending order of frequency in $b$.
Then 
$\Pr(\land_{i \in [l], j \in [k_i]} \lnot A_{i, j} \|{\B})$
is minimized over all atoms $A_{i, j}$ when,
$A_{i, j}$ is $t_{p_i}[S] = s_b^j$, for all $i \in [l]$ and all $j \in [k_i]$, 
where $p_0, p_1, \dots, p_{l - 1} \in P_b$ are distinct.
Consequently, the minimum probability is given by:
\begin{eqnarray}
&\prod_{i \in [l]} \frac{n_b - i - {\sum_{j \in [k_i]} n_b(s_b^j)}}{n_b - i}&
%\label{eqn:closed-form}
\end{eqnarray}
\end{lemma}

Note that $l \leq k$ and $k = \sum_{i \in [l]} k_i$ since each atom involves at exactly one person.
So the question of minimizing
$\Pr(\land_{i \in [k]} \lnot A_i \| {\B})$ over all atoms $A_i$ that mention only tuples in $b$
becomes one of minimizing
$\prod_{i \in [l]} \frac{n_b - i - \sum_{j \in [k_i]} n_b(s_b^j)}{n_b - i}$
over all $l \leq k$ and all $k_0, \dots, k_{l - 1}$ such that $\sum_{i \in [l]} k_i = k$.

\begin{algorithm}[tb]
\caption{: $\MinimizeWithinBucket(b, i, \hat{k}_i, \hat{k})$}
\label{alg:withinbucket} 
\begin{algorithmic}[1] 
{\footnotesize
  \INPUT \textit{$b$ is the bucket under consideration}
  \INPUT \textit{$i$ is the index of the next person $p_i$ for which $k_i$ (i.e., the number of atoms involving person $p_i$) is to be determined (initially $0$)}
  \INPUT \textit{$\hat{k}_i$ is the the upper bound for $k_i$ (initially $k$)}
  \INPUT \textit{$\hat{k}$ is the number of atoms for which the people involved have yet to be been determined (initially $k$)}
  \STATE $p_{\min} \from 1$
  \FOR{$k_i = 1, 2, \ldots, \min(\hat{k}_i, \hat{k})$}
    \STATE $p \from \MinimizeWithinBucket(b, i + 1, k_i, \hat{k} - k_i)$
    \STATE $p \from \frac{n_b - i - \sum_{j \in [k_i]} n_b(s_b^j)}{n_b - i} \times p$
    \STATE $p_{\min} \from \min(p_{\min}, p)$ 
  \ENDFOR 
  \STATE {\bf return} $p_{\min}$
}
\end{algorithmic} 
\end{algorithm}

This can easily be done using Algorithm $\ref{alg:withinbucket}$.
Thus, calling $\MinimizeWithinBucket(b, 0, k, k)$ minimizes $\Pr(\land_{i \in [k]} \lnot A_i \| \phi_{\B})$ over all atoms $A_i$ that involve people with tuples in bucket $b$.
It is easy to modify the algorithm to remember the minimizing values of $k_0, \dots, k_{l - 1}$, 
and thus we can even reconstruct the set of minimizing atoms according to Lemma $\ref{lem:impwithin}$.

\textbf{Algorithm complexity. }
Note that the parameters of $\MinimizeWithinBucket$ are bounded.  That is, for every recursive call $\MinimizeWithinBucket(b, i, k_i, \hat{k})$ that occurs inside the initial call to $\MinimizeWithinBucket(b, 0, k, k)$,
parameter $b$ does not change, and
parameters $i, \hat{k}_i, \hat{k}$ are all bounded by $k$ (i.e., the number of implications we allow the attacker to know).
So we can easily turn this into an $O(k^3)$ time and space algorithm using dynamic programming.

\SubSubSection{Minimizing Formula (\ref{eqn:tomin}) within one bucket}\label{sec:alg:2}

Let us now minimize $\frac{\Pr(\lnot A \land (\land_{i \in [k]} \lnot A_i) \| {\B})}{\Pr(A \| {\B})}$ over all $k + 1$ atoms $A$ and $A_i$, for $i \in [k]$, that only mention tuples in bucket $b$.
Clearly any $A, A_i$ that simultaneously minimize the numerator and maximize the denominator will work.
We know that $\MinimizeWithinBucket(b, 0, k + 1, k + 1)$ will minimize the numerator.  According to Lemma \ref{lem:impwithin}, at least one of these minimal $k + 1$ atoms mention the most frequent sensitive value.  So, taking this atom to be $A$, we maximize the denominator as well.  Thus, the minimum value is
$$\MinimizeWithinBucket(b, 0, k + 1, k + 1) \times \frac{n_b}{n_b(s^0_b)}.$$

\SubSubSection{Minimizing Formula (\ref{eqn:tomin}) over all buckets}\label{sec:alg:3}

We look again at minimizing $\frac{\Pr(\lnot A \land (\land_{i \in [k]} \lnot A_i) \| {\B})}{\Pr(A \| {\B})}$, 
except this time, we allow $A$ and $A_i$ for $i \in [k]$ to mention tuples in possibly different buckets.
To do this, we make use of the independence between buckets.
Suppose that the $k + 1$ minimizing atoms (including $A$) are such that $k_i$ of them mention tuples in 
bucket $b_i$, for each $i \in [l]$ for some $l \leq k + 1$.
Let $b_j$ be the bucket containing the tuple mentioned by $A$.
Then, since the permutation of sensitive values for each bucket was picked independently, we can compute the minimum as
$$
\frac{n_{b_j}}{n_{b_j}(s^0_{b_j})} \times \prod_{i \in [l]} \MinimizeWithinBucket(b_i, 0, k_i, k_i).
$$
So we need to minimize the above for all choices of $l \leq k + 1$, $j$, and $k_0, k_1, \dots, k_{l - 1}$ 
(which we can assume without loss of generality to be in descending order).
Assuming buckets in $\B$ are labeled as $b_0, b_1, b_2, \dots$, this is done by the $\MinimizeAcrossBuckets$.

\begin{algorithm}[tb]
\caption{: $\MinimizeAcrossBuckets(i, h_i, a)$}
\label{alg:acrossbuckets} 
\begin{algorithmic}[1] 
{\footnotesize
  \INPUT \textit{$i$ is the current bucket $b_i$ (initially $0$)}
  \INPUT \textit{$h_i$ is number of atoms $A_j, j \in [k]$ that we have yet to determine (initially $k$)}
  \INPUT \textit{$a$ is a flag representing whether atom $A$ involves a person in an earlier bucket $b_j$, $j < i$ (initially $\false$)}
  \STATE $r_{min} \from \infty$
  \IF{$i = |\B|$}
    \STATE \textit{// Finished all buckets}
    \STATE {\bf return} $r_{min}$
  \ENDIF
  \FOR{$h_{i+1} = 0, 1, 2, \ldots, h_i$}
    \STATE $u \from \MinimizeWithinBucket(b_i, 0, h_{i + 1}, h_{i + 1})$
    \STATE $x \from \MinimizeAcrossBuckets(i + 1, h_i - h_{i + 1}, \true)$
    \IF{$a = \false$}
      \STATE \textit{// Atom $A$ does not involve an earlier bucket $b_j$, $j < i$}
      \STATE \textit{// So either $A$ involves $b_i$...}
      \STATE $v \from \MinimizeWithinBucket(b_i, 0, h_{i + 1} + 1, h_{i + 1}+1)$
      \STATE $r_{\min} \from \min(r_{\min}, v \times x \times \frac{n_{b_i}}{n_{b_i}(s_{b_i}^0)})$
      \STATE \textit{// ... or else $A$ involves a later bucket $b_j$, $j > i$}
      \STATE $r_{\min} \from \min(r_{\min}, u \times \MinimizeAcrossBuckets(i + 1, h_i - h_{i + 1}, \false))$
    \ELSE
      \STATE \textit{// Atom $A$ involves an earlier bucket $b_j$, $j < i$}
      \STATE $r_{\min} \from \min(r_{\min}, u \times x)$
    \ENDIF
  \ENDFOR 
  \STATE {\bf return} $r_{min}$
}
\end{algorithmic} 
\end{algorithm}

So $\MinimizeAcrossBuckets(0, k, \true)$ minimizes $\frac{\Pr(\lnot A \land (\land_{i \in [k]} \lnot A_i) \| {\B})}{\Pr(A \| {\B})}$ 
over all atoms $A, A_i$, $i \in [k]$. It is easy to modify the algorithm to remember the $i$'s and $h_i$'s, and hence reconstruct the minimizing atoms.

\textbf{Algorithm complexity. }
Note that the parameters of $\MinimizeAcrossBuckets$ are bounded.  That is, for every recursive call to $\MinimizeAcrossBuckets(i, h_i, a)$ that occurs inside the initial call to $\MinimizeAcrossBuckets(0, k, \true)$,
parameter $i$ is bounded by the number of buckets, 
parameter $k_i$ is bounded by the total number of implications $k$, and
$a$ is either $\true$ or $\false$.
Thus, assuming that we first memoize (i.e., precompute all possible calls to) $\MinimizeWithinBucket$
(which we can do in time $O(|\B| \times k^3)$),
we can modify the $\MinimizeAcrossBuckets$ algorithm using dynamic programming to take an additional $O(|\B| \times k)$time and space.
So the whole algorithm can be made to run in $O(|\B| \times k^3)$time and space.

Incidentally, if one had two bucketizations $\B$ and $\B^*$ that differed only in that 
$\B^*$ was the result of removing some buckets from $\B$ and adding $x$ new buckets to $\B$, 
then, after we run the algorithm for $\B$, we memoize $\MinimizeWithinBucket$ for the $x$ new buckets;
so the incremental cost of running the algorithm for $\B^*$ is $O(|\B^*| \times k + x \times k^3)$-time.
Moreover, if one knew in advance which buckets were going to be removed, one could order the buckets $b_0, b_1, \dots$ 
appropriately to reuse much of the memoization of $\MinimizeAcrossBuckets$ as well.

%\subsection{Maximizing $\Pr(A \| {\B} \land (\land_{i \in [k]} A_i \to A))$ for many buckets}

\SubSection{Finding a safe bucketization}\label{sec:imp:monotonicity}

Armed with a method to compute the maximum disclosure, we now show how to efficiently find a ``minimally sanitized'' bucketization for which maximum disclosure is below a given threshold.
Intuitively, we would like a minimal sanitization in order to preserve the utility of the published data.
Let us be more concrete about the notion of minimal sanitization.
Given a table, consider the set of bucketizations of this table.
We impose a partial ordering $\preceq$ on this set of bucketizations where $\B \preceq \B'$ if and only if 
every bucket in $\B'$ is the union of one of more buckets in $\B$.
Thus the bucketization $\B_\top$ that has all the tuples in one bucket is the unique top element of this partial order,
and the bucketization $\B_\bot$ that has one tuple per bucket is the unique bottom element of this partial order.
Our notion of a ``minimally sanitized'' bucketization is one that is as low as possible in the partial order (i.e., as close to $\B_\bot$) while still having maximum disclosure lower than a specified threshold.
\begin{definition}[$(c, k)$-safety]
Given a threshold $c \in [0, 1]$, we say that $\B$ is \emph{a $(c, k)$-safe bucketization} if
the maximum disclosure of $\B$ with respect to $\Lb{k}$ is less than $c$.
\end{definition}
If the maximum disclosure is \emph{monotonic} with respect to the partial ordering $\preceq$, then finding a $\preceq$-minimal $(c, k)$-safe bucketization can be done in time logarithmic in the height of the bucketization lattice (which is at most the number of tuples in the table) by doing a binary search.  
The following theorem says that we do indeed have monotonicity.
\begin{theorem}[Monotonicity]\label{thm:monotonicity}
Let $\B$ and $\B'$ be bucketizations such that $\B \preceq \B'$.
Then the maximum disclosure of $\B$ is at least as high as the maximum disclosure of $\B'$ with respect to $\Lb{k}$.
\end{theorem}
Another approach is to find \emph{all} $\preceq$-minimal $(c, k)$-safe bucketizations, and return the one that maximizes a specified utility function.
The monotonicity property allows us to make use of existing algorithms for efficient itemset mining \cite{vldb:AS94}, $k$-anonymity \cite{bayardoA05:optimalkAnon, lefevreDR05:incognito} and $\ell$-diversity \cite{machanavajjhalaGKV06:l-diversity}.\footnote{While these algorithms typically have worst-case exponential running time in the height of the bucketization lattice, they have been shown to run fast in practice.}
For example, we can modify the Incognito \cite{lefevreDR05:incognito} algorithm, which finds all the $\preceq$-minimal $k$-anonymous bucketizations, by simply replacing the check for $k$-anonymity with the check for $(c, k)$-safety from Section \ref{sec:imp:algorithm}.  We can thus find the bucketization that maximizes a given utility function subject to the constraint that the bucketization be $(c, k)$-safe.

%% file: experiments.tex
\Section{Experiments}\label{sec:experiments}

\begin{figure*}[t]
\begin{minipage}{3.2in}
\centering
\includegraphics[angle=-90,width=2.8in]{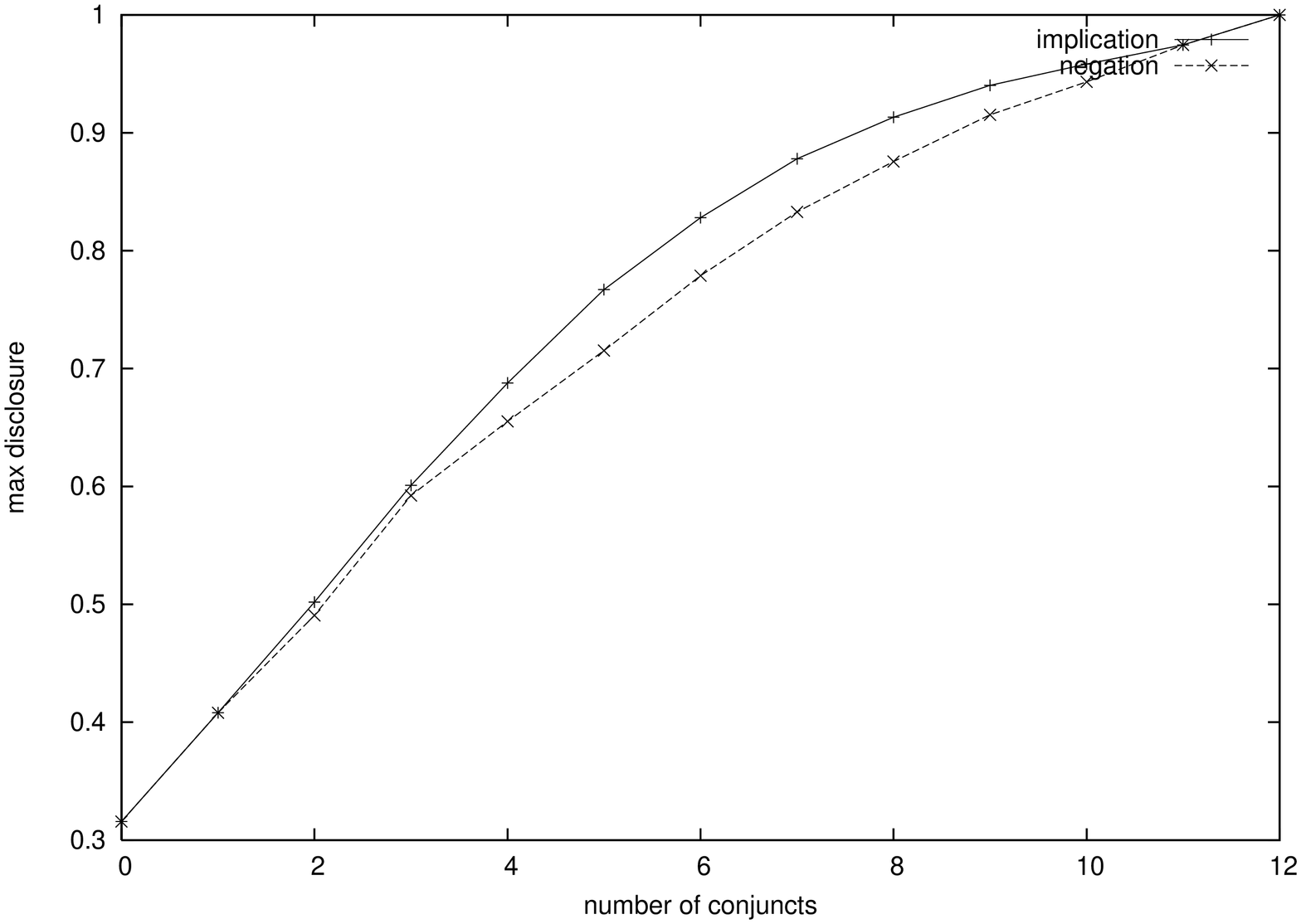}
\end{minipage}
\hfill
\begin{minipage}{3.2in}
\centering
\includegraphics[angle=-90,width=2.8in]{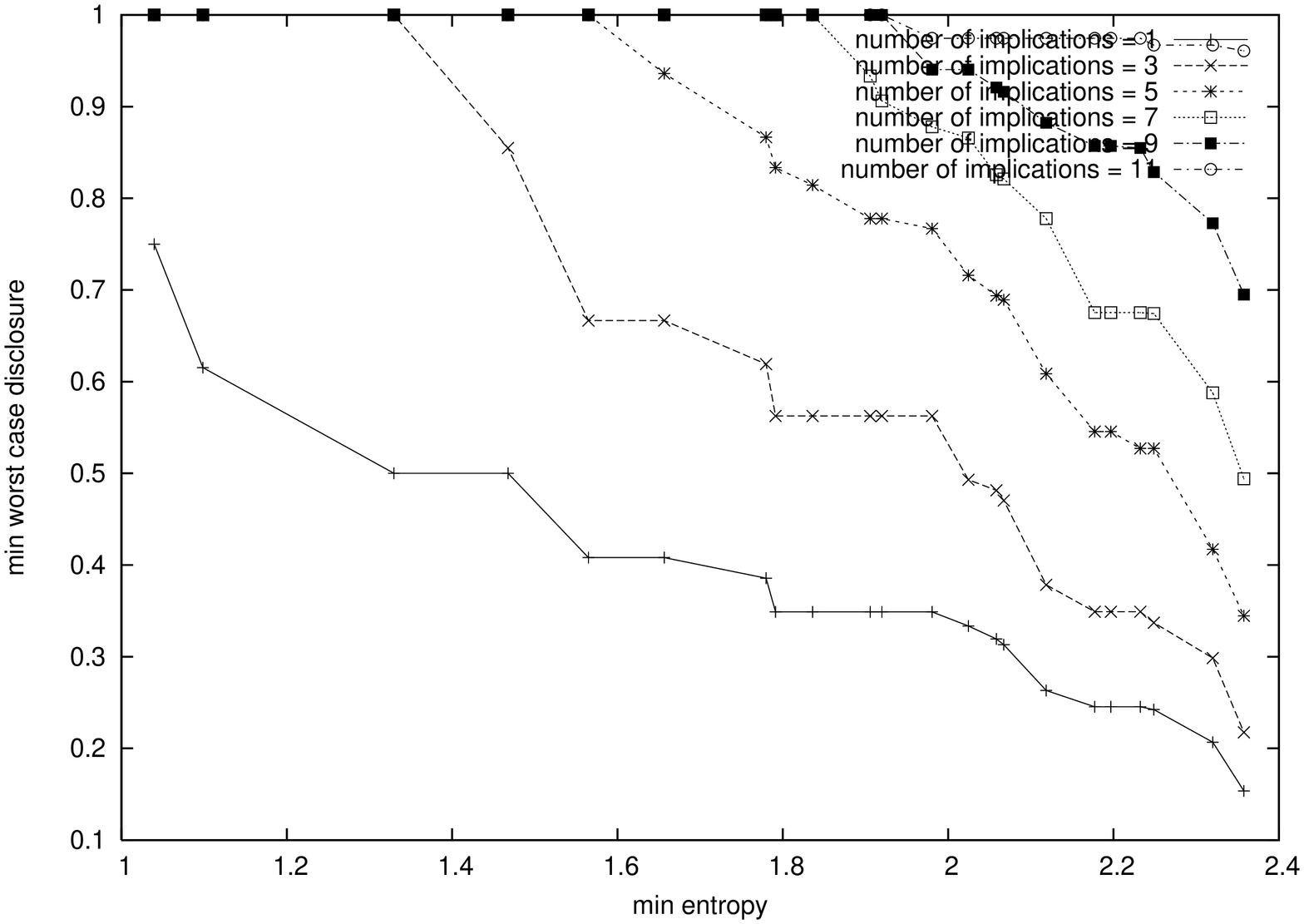}
\end{minipage}
\vspace {-0.5em} 
\centerline{
\parbox{3.5in}{\caption{\label{fig:lpa}\small{Disclosure vs \# pieces of background knowledge}}} 
\hfill
\parbox{3.5in}{\caption{\label{fig:hp}\small{Entropy vs Maximum Disclosure Risk}}}}
\end{figure*}

In this section, we present a case-study of our framework for worst-case disclosure
using the Adult Database from the UCI Machine Learning Repository \cite{uciml:repository}.  
We only consider the projection of the Adult
Database onto five attributes -- Age, Marital Status, Race, Gender and
Occupation.  The dataset has 45,222 tuples after removing tuples with missing
values.  We treat Occupation as the sensitive attribute; its domain consists
of fourteen values.  We use pre-defined generalization hierarchies for the
attributes similar to the ones used in \cite{lefevreDR05:incognito}. Age can
be generalized to six levels (unsuppressed, generalized to intervals of size 5, 10,
20, 40, or completely suppressed), Marital Status can be generalized to three
levels, and Race and Gender can each either be left as is or be completely
suppressed. 
We consider all the possible anonymized tables using those generalizations.

We computed the maximum disclosure for $k$ pieces of background knowledge, for $k$ ranging from $0$
(i.e., no background knowledge) to $12$
(since we know that maximum disclosure certainly reaches $1$ at $k = 13$ because there are only fourteen possible sensitive values).
Figure $\ref{fig:lpa}$ plots, for one
anonymized table, the number of pieces of knowledge available to
an adversary against the maximum disclosure for both negated atoms
($\ell$-diversity) and basic implications.  
In the anonymized table used, all
the attributes other than Age were suppressed and the Age attribute was
generalized to intervals of size $20$.  
The solid line corresponds to implication statements and the dotted line
corresponds to negated atoms.  This graph agrees with our earlier observation
that implication-type background knowledge subsumes negation; the maximum
disclosure for $k$ negated atoms is always smaller than the
maximum disclosure for $k$ implications. However, note that, for a given $k$,
the difference between the maximum disclosure for negated atoms
and for basic implications is not too large.  This means that an anonymized table which tolerates maximum disclosure due to $k$ negated atoms need not be anonymized much further to
defend against $k$ implications.

Intuitively, if all the buckets in a table have a nearly uniform distribution,
then the maximum disclosure should be lower, but the exact
relationship is not obvious. To get a better picture, we performed the following
experiment. We fixed a value $k$ for the number of pieces of information. For
every entropy value $h$, we looked at all tables ${\cal T}(h)$ for which the minimum
entropy of the sensitive attribute over all buckets was equal to $h$. Amongst ${\cal
T}(h)$ we found the table $T(h)$ with the least maximum disclosure for
$k$ implications. Let the worst case disclosure for $T(h)$ given $k$
pieces of knowledge be denoted by $w(T(h), k)$. We plotted $h$ versus $w(T(h),
k)$ for $k = 1, 3, 5, 7, 9, 11$ in Figure \ref{fig:hp}. We see a
behavior which matches our intuition.  For a given $k$, the disclosure
risk monotonically decreases with increase in $h$. This is because increasing
$h$ means that we are looking at tables with more and more entropy in their
buckets (and, consequently, less skew). We plotted an analogous graph (which we do not show here) for negation
statements and observed very similar behavior.

%% file: relatedwork.tex
\Section{Related Work}\label{sec:relatedwork}
Many metrics have been proposed to quantify privacy guarantees in publishing
publishing anonymized data-sets.
`Perfect privacy' \cite{deutschP05:privacy, miklauS04:disclosure} guarantees that
published data does not disclose any information about the sensitive
data. However, checking whether a conjunctive query discloses any information
about the answer to another conjunctive query is shown to be very hard
($\Pi^p_2$-complete \cite{miklauS04:disclosure}). Subsequent work showed that
checking for perfect privacy can be done efficiently for many subclasses of
conjunctive queries \cite{machanavajjhalaG06:perfect}. Perfect privacy places
very strong restrictions on the types of queries that can be answered
\cite{miklauS04:disclosure} (in particular, aggregate statistics cannot be
published).  Less restrictive privacy definitions based on asymptotic
conditional probabilities \cite{dalviMS05:asymptotic} and certain answers
\cite{stoffelS05:provable} have been proposed. Statistical databases allow
answering aggregates over sensitive values without disclosing the exact value
\cite{adamW89:security}. De-identification, like $k$-anonymity
\cite{samarati01:microdata,sweeney02:kAnon} and ``blending in a crowd''
\cite{chawlaDMSW05:towardPrivacy}, ensures that an individual cannot be
associated with a unique tuple in an anonymized table.
However, under both of those definitions, sensitive information can be disclosed if
groups are homogeneous. 

Background knowledge can lead to disclosure of sensitive information.
Su et al. \cite{suO:FDMVD} and Yang et al. \cite{yangL04:secureXML} limit
disclosure when functional dependencies in the data are known to the data
publisher upfront.
The notion of
$\ell$-diversity \cite{machanavajjhalaGKV06:l-diversity} guards against
limited amounts of background knowledge unknown to the data publisher.
Farkas et al. \cite{farkasJ:survey} provide a survey of indirect data
disclosure via inference channels.

There are several approaches to anonymizing a dataset to ensure privacy. These
include generalizations \cite{bayardoA05:optimalkAnon, lefevreDR05:incognito,
samaratiS98:protecting}, cell and tuple suppression
\cite{cox80:suppression,samaratiS98:protecting}, adding noise
\cite{adamW89:security, agrawalS00:PPDM, chawlaDMSW05:towardPrivacy,
sashaGS03:limiting}, publishing marginals that satisfy a safety range
\cite{dobra02:thesis}, and data swapping \cite{daleniusR82:swapping}, where attributes are swapped between tuples so that certain marginal totals are preserved. 
Queries can be posed online and
the answers audited \cite{kenthapadiMN05:simulatable} or perturbed
\cite{dinurN03:sulq}. Not all approaches guarantee privacy. For example,
spectral techniques can separate much of the noise from the data if the noise is uncorrelated with the data \cite{huangDC04:deriving, karguptaDWS03:randomPerturbation}.
Anatomy \cite{xiao06:anatomy} is a recently proposed anonymization technique that corresponds exactly to the notion of bucketization that we use in this paper.
When the attacker knows full identification information, then generalization provides no more privacy than bucketization.  
However, we recommend generalizing the attributes before publishing the data since this will prevent attackers that do not already have full identification information from reidentifying individuals via linking attacks \cite{sweeney02:kAnon}.
In many cases, the fact that a particular individual is in the table is considered sensitive information \cite{chawlaDMSW05:towardPrivacy}.

The utility of data that has been altered to preserve privacy has often been
studied for specific future uses of the data.
Work has been done on preserving association rules
while adding noise \cite{sashaGS03:limiting}; reconstructing distributions of continuous variables after adding noise with a known distribution \cite{agrawalS00:PPDM,agrawalA01:quantificationPPDM}; reconstructing data clusters after perturbing numeric attributes \cite{chawlaDMSW05:towardPrivacy}; and maximizing decision tree accuracy while anonymizing data \cite{iyengar02:transform, wang05:template}.
There have also been some negative
results for utility. Publishing a single $k$-anonymous table can suffer from the curse of dimensionality \cite{aggarwal05:kcurse} - large
portions of the data need to be suppressed to ensure
privacy. Subsequent work \cite{kiferG06:injecting} shows how to publish several
tables instead of a single one to combat this.

%% file: conclusions.tex
\Section{Conclusions}\label{sec:conclusions}

In this paper, we initiate a formal study of the worst-case disclosure 
with background knowledge.  Our analysis does not assume that we
are aware of the exact background knowledge possessed by the attacker.  We
only assume bounds on the the attacker's background knowledge in terms of the
number of basic units of knowledge that the attacker possesses.  We propose
basic implications as an expressive choice for these units
of knowledge.  Although computing the probability of a specific disclosure from
 a given set of $k$ basic implications is intractable, we show how to
efficiently determine the worst-case over all sets of $k$ basic implications.
In addition, we show how to search for a bucketization that is robust (to a desired threshold $c$) against any $k$ basic implications by combining our check for $(c, k)$-safety with existing lattice-search algorithms.  
Finally, we demonstrate that, in practice, $\ell$-diversity has similar maximum
disclosure to our notion of $(c, k)$-safety, which guards against a richer class of background knowledge.

Since we chose basic implications as our units of knowledge, our algorithms
will clearly yield very conservative bucketizations if we try to protect against an
attacker who knows information that can only be expressed using a large number
of basic implications. 
One way to reduce the number of basic units required is to add more powerful atoms to our existing language.  
Finding the right language for basic units of knowledge is an important
direction of future work.

Other directions for future work include extending our framework for
probabilistic background knowledge, studying cost-based disclosure (since it
was observed in \cite{machanavajjhalaGKV06:l-diversity} that not all
disclosures are equally bad), and extending our results to other forms
of anonymization, such as data-swapping and collections of anonymized marginals
\cite{kiferG06:injecting}.

\textbf{Acknowledgments.}
This work was supported by the National Science Foundation under Grants IIS-0541507, IIS-0636259, CNS-0627680, IIS-0534064, and ITR-0325453; by the DoD Multidisciplinary University Research Initiative (MURI) program administered by the ONR under grants N00014-01-1-0795 and N00014-04-1-0725; by AFOSR under grant FA9550-05-1-0055; by a Sloan Foundation Fellowship; and by gifts from Yahoo!~and Microsoft. Any opinions, findings, conclusions, or recommendations expressed in this material are those of the authors and do not necessarily reflect the views of the sponsors.

%% file: main.bbl
\begin{thebibliography}{10}

\bibitem{adamW89:security}
N.~R. Adam and J.~C. Wortmann.
\newblock Security-control methods for statistical databases: a comparative
  study.
\newblock {\em ACM Comput. Surv.}, 21(4):515--556, 1989.

\bibitem{aggarwal05:kcurse}
Charu~C. Aggarwal.
\newblock On k-anonymity and the curse of dimensionality.
\newblock In {\em VLDB}, pages 901--909, 2005.

\bibitem{agrawalA01:quantificationPPDM}
D.~Agrawal and C.~C. Aggarwal.
\newblock On the design and quantification of privacy preserving data mining
  algorithms.
\newblock In {\em PODS}, 2001.

\bibitem{vldb:AS94}
R.~Agrawal and R.~Srikant.
\newblock Fast algorithms for mining association rules in large databases.
\newblock In {\em VLDB}, 1994.

\bibitem{agrawalS00:PPDM}
R.~Agrawal and R.~Srikant.
\newblock Privacy preserving data mining.
\newblock In {\em SIGMOD}, 2000.

\bibitem{bacchusGHK96:randomworld}
F.~Bacchus, A.~J. Grove, J.~Y. Halpern, and D.~Koller.
\newblock From statistical knowledge bases to degrees of belief.
\newblock {\em A.I.}, 87(1-2), 1996.

\bibitem{bayardoA05:optimalkAnon}
R.~J. Bayardo and R.~Agrawal.
\newblock Data privacy through pptimal k-anonymization.
\newblock In {\em ICDE}, 2005.

\bibitem{chawlaDMSW05:towardPrivacy}
S.~Chawla, C.~Dwork, F.~McSherry, A.~Smith, and H.~Wee.
\newblock Toward privacy in public databases.
\newblock In {\em TCC}, 2005.

\bibitem{cox80:suppression}
L.~H. Cox.
\newblock Suppression, methodology and statistical disclosure control.
\newblock {\em Journal of the American Statistical Association}, 75, 1980.

\bibitem{daleniusR82:swapping}
T.~Dalenius and S.~Reiss.
\newblock Data swapping: a technique for disclosure control.
\newblock {\em Journal of Statistical Planning and Inference}, 6, 1982.

\bibitem{dalviMS05:asymptotic}
N.~Dalvi, G.~Miklau, and D.~Suciu.
\newblock Asymptotic conditional probabilities for conjunctive queries.
\newblock In {\em ICDT}, 2005.

\bibitem{deutschP05:privacy}
A.~Deutsch and Y.~Papakonstantinou.
\newblock Privacy in database publishing.
\newblock In {\em ICDT}, 2005.

\bibitem{dinurN03:sulq}
I.~Dinur and K.~Nissim.
\newblock Revealing information while preserving privacy.
\newblock In {\em PODS}, pages 202--210, 2003.

\bibitem{dobra02:thesis}
A.~Dobra.
\newblock {\em Statistical tools for disclosure limitation in multiway
  contingency tables}.
\newblock PhD thesis, Carnegie Mellon University, 2002.

\bibitem{sashaGS03:limiting}
A.~Evfimievski, J.~Gehrke, and R.~Srikant.
\newblock Limiting privacy breaches in privacy preserving data mining.
\newblock In {\em PODS}, 2003.

\bibitem{farkasJ:survey}
C.~Farkas and S.~Jajodia.
\newblock The inference problem: a survey.
\newblock {\em SIGKDD Explor. Newsl.}, 4(2), 2002.

\bibitem{huangDC04:deriving}
Z.~Huang, W.~Du, and B.~Chen.
\newblock Deriving private information from randomized data.
\newblock In {\em SIGMOD}, 2004.

\bibitem{iyengar02:transform}
Vijay~S. Iyengar.
\newblock Transforming data to satisfy privacy constraints.
\newblock In {\em KDD}, pages 279--288, 2002.

\bibitem{karguptaDWS03:randomPerturbation}
H.~Kargupta, S.~Datta, Q.~Wang, and K.~Sivakumar.
\newblock On the privacy preserving properties of random data perturbation
  techniques.
\newblock In {\em ICDM}, pages 99--106, 2003.

\bibitem{kenthapadiMN05:simulatable}
K.~Kenthapadi, N.~Mishra, and K.~Nissim.
\newblock Simulatable auditing.
\newblock In {\em PODS}, 2005.

\bibitem{kiferG06:injecting}
Daniel Kifer and Johannes Gehrke.
\newblock Injecting utility into anonymized datasets.
\newblock In {\em SIGMOD}, 2006.

\bibitem{lefevreDR05:incognito}
K.~LeFevre, D.~DeWitt, and R.~Ramakrishnan.
\newblock Incognito: Efficient fulldomain k-anonymity.
\newblock In {\em SIGMOD}, 2005.

\bibitem{machanavajjhalaG06:perfect}
A.~Machanavajjhala and J.~Gehrke.
\newblock On the efficiency of checking perfect privacy.
\newblock In {\em PODS}, 2006.

\bibitem{machanavajjhalaGKV06:l-diversity}
A.~Machanavajjhala, J.~Gehrke, D.~Kifer, and M.~Venkitasubramaniam.
\newblock $\ell$-diversity: Privacy beyond $k$-anonymity.
\newblock In {\em ICDE}, 2006.

\bibitem{martinKMGH06:worstCaseTR}
D.~Martin, D.~Kifer, A.~Machanavajjhala, J.~Gehrke, and J.~Halpern.
\newblock Worst-case background knowledge in privacy.
\newblock Technical report, Cornell University, 2006.

\bibitem{miklauS04:disclosure}
G.~Miklau and D.~Suciu.
\newblock A formal analysis of information disclosure in data exchange.
\newblock In {\em SIGMOD}, 2004.

\bibitem{uciml:repository}
U.C. Irvine Machine~Learning Repository.
\newblock http://www.ics.uci.edu/~mlearn/mlrepository.html.

\bibitem{samarati01:microdata}
P.~Samarati.
\newblock Protecting respondents' identities in microdata release.
\newblock In {\em IEEE Transactions on Knowledge and Data Engineering}, 2001.

\bibitem{samaratiS98:protecting}
P.~Samarati and L.~Sweeney.
\newblock Protecting privacy when disclosing information: k-anonymity and its
  enforcement through generalization and suppression.
\newblock Technical report, CMU, SRI, 1998.

\bibitem{stoffelS05:provable}
K.~Stoffel and M.~Studer.
\newblock Provable data privacy.
\newblock In {\em DEXA}, 2005.

\bibitem{suO:FDMVD}
T.~Su and G.~Ozsoyoglu.
\newblock Controlling fd and mvd inferences in multilevel relational database
  systems.
\newblock {\em IEEE TKDE}, 3(4), 1991.

\bibitem{sweeney02:kAnon}
L.~Sweeney.
\newblock k-anonymity: a model for protecting privacy.
\newblock {\em International Journal on Uncertainty, Fuzziness and
  Knowledge-based Systems}, 10(5):557--570, 2002.

\bibitem{wang05:template}
K.~Wang, B.~C.~M. Fung, and P.~S. Yu.
\newblock Template-based privacy preservation in classification problems.
\newblock In {\em ICDM}, November 2005.

\bibitem{xiao06:anatomy}
X.~Xiao and Y.~Tao.
\newblock Anatomy: Simple and effective privacy preservation.
\newblock In {\em VLDB}, 2006.

\bibitem{yangL04:secureXML}
X.~Yang and C.~Li.
\newblock Secure xml publishing without information leakage in the presence of
  data inference.
\newblock In {\em VLDB}, pages 96--107, 2004.

\end{thebibliography}
